\documentclass[conference]{IEEEtran}
\usepackage{cite}
\usepackage[english]{babel}
\usepackage{amsmath,amssymb,amsfonts}
\usepackage{algorithmic}
\usepackage{listing}
\usepackage{booktabs}
\usepackage[frozencache]{minted}
\usepackage{graphicx}
\usepackage{hyperref}
\usepackage{textcomp}
\usepackage{braket}
\usepackage{subfig}
\usepackage{xcolor}
\def\BibTeX{{\rm B\kern-.05em{\sc i\kern-.025em b}\kern-.08em
    T\kern-.1667em\lower.7ex\hbox{E}\kern-.125emX}}

\DeclareUnicodeCharacter{221A}{$\sqrt{\phantom{2}}\mkern-9mu$}
\DeclareUnicodeCharacter{2245}{$\cong$}

\setminted[ipython]{frame=lines}
\setminted[ketcon]{xleftmargin=2em,frame=lines,linenos}
\setminted[ket]{xleftmargin=2em,frame=lines,linenos}

\hypersetup{hidelinks}

\begin{document}

\title{Programming with Quantum Mechanics}

\author{\IEEEauthorblockN{Evandro C. R. da Rosa}
    \IEEEauthorblockA{\textit{Quantuloop}}
    \and
    \IEEEauthorblockN{Claudio Lima}
    \IEEEauthorblockA{\textit{Quantuloop}}
}

\maketitle

\begin{abstract}
    Quantum computing is an emerging paradigm that opens a new era for exponential computational speedup. Still, quantum computers have yet to be ready for commercial use. However, it is essential to train and qualify today the workforce that will develop quantum acceleration solutions to get the quantum advantage in the future. This tutorial gives a broad view of quantum computing, abstracting most of the mathematical formalism and proposing a hands-on with the quantum programming language Ket. The target audience is undergraduate and graduate students starting in quantum computing -- no prerequisites for following this tutorial.
\end{abstract}


\tableofcontents

\section{Introduction}

Quantum computing opens up new ways for processing information, solving in a few minutes problems that would take decades for a classical computer to find the solution. This claim was known in theory for a long time. Now we begin to see the first quantum computers that outperform classical ones. This tutorial walks through some quantum mechanics concepts with the eyes of quantum computing, demonstrating how to use them to program a quantum application.

The idea of quantum computers dates back to the 80s when Richard Feynman proposed a quantum computer as a universal quantum simulator. His motivation was the difficulty of simulating quantum systems in classical computers, a problem with a time complexity that grows exponentially with the number of variables. The proof that quantum computers can significantly increase processing power came in the late 90s with the work of Peter Shor. His integer factorization algorithm, known as Shor's algorithm, reveals how to solve a problem that tokens exponential time in a classical computer in polynomial time with the help of a quantum computer.

Shor's algorithm boosted the development of quantum computers and uplifted the creation of post-quantum cryptography. A research field to find quantum-resistant alternatives for today's standard public-key cryptography algorithms since Shor's algorithm breaks all of them. While this may sound alarming, the industry still lacks powerful quantum computers to break a standard cryptography scheme. Moreover, NIST is working to standardize new quantum-resistant asymmetric encryption algorithms.

Quantum computers can speed up several processes, including but not limited to optimization, logistics, machine learning, and quantum chemistry simulation. However, we are in the Noisy Intermediate-Scale Quantum (NISQ) era, with quantum computers featuring few qubits very susceptible to noise that limits the complexity of quantum executions. Although, we went further than was expected 20 years ago, even reaching the quantum advantage milestone with quantum computers outperforming classical ones for some tasks. The task, in this case, is not solving any real-world problem. It is just a trial projected specifically for the quantum advantage demonstration. However, we are not far from large-scale fault-tolerant quantum computers, being on the roadmap of many companies to deliver them by the end of the decade (until 2030).

Quantum engineers are a highly in-demand workforce today, even though we have not yet unlocked the full potential of quantum computing. We expect this demand to grow in the foreseeable future. A new domain that comes with the growth of quantum technologies is the quantum developer, professionals who adapt solutions to take advantage of quantum computers and programming quantum applications.

Adapting and developing quantum algorithms is not a straightforward process. Although, quantum programming is not as hard as one may think. It is much like classical programming. This tutorial will discuss the main characteristics of quantum computing, demonstrating how to express it in the quantum programming language \emph{Ket}. We will present a broad view of the topic, which may help one take the first steps in quantum development.

\emph{Ket} is an embedded programming language that introduces the ease of Python to quantum programming, letting anyone quickly prototype and test quantum applications. We expect the reader following this tutorial to be familiar with Python programming and Jupyter notebook. We will introduce quantum programming focusing on digital qubit-based quantum computing, a general-purpose quantum computation model used by several companies, including IBM, Google, and Microsoft.

\section{Setup}

\begin{figure}[h]
    \centering
    \includegraphics[width=\linewidth]{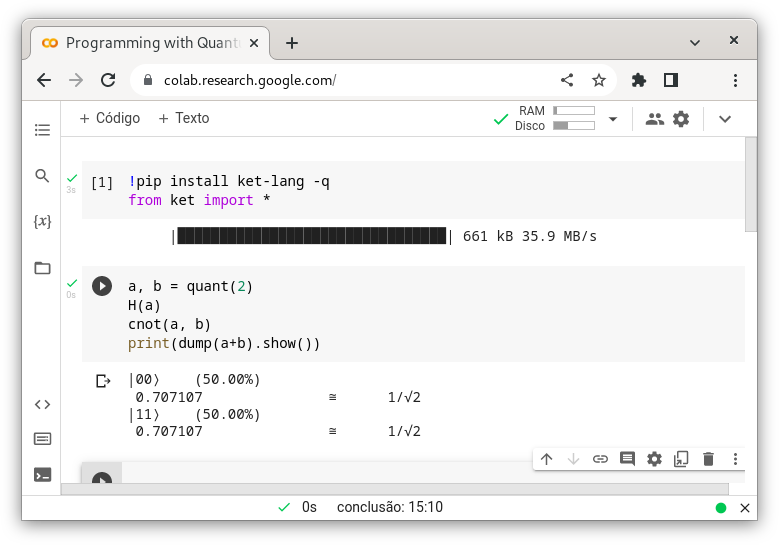}
    \caption{Running Ket on Google Colab.}\label{fig:colab}
\end{figure}

We will present some examples that one can try on their computer. There are several ways one can execute a Ket program. Regardless, using Google Colab is the most straightforward approach, allowing one to create and run Ket applications on any computer quickly. However, feel free to use any method. Get started by going to the Colab website, creating a new notebook, and running the command presented in Listing \ref{code:install}. See Ket's project website\footnote{\url{https://quantumket.org}} for more ways to get Ket. Try running the example from Listing \ref{code:bell} to test if it works. It should output similarly as Fig. \ref{fig:colab}.

\begin{listing}[h]
    \centering
    \begin{minted}{ipython}
!pip install ket-lang qutip -q
from ket import *
    \end{minted}
    \caption{Installing and importing Ket on a Jupyter notebook.}\label{code:install}
\end{listing}

Listing \ref{code:bell} allocates two qubits, uses a Hadamard gate to create superposition, a controlled gate to entangle both qubits, and prints the quantum state on the screen. Do not worry if some of those terms seem unfamiliar; this tutorial will explain all those concepts, presenting how to use them in quantum programming.

\begin{listing}[h]
    \centering
    \begin{minted}{ketcon}
>>> from ket import *
>>> a, b = quant(2)
>>> H(a)
>>> cnot(a, b)
>>> print(dump(a+b).show())
|00⟩    (50.00%)
 0.707107               ≅      1/√2
|11⟩    (50.00%)
 0.707107               ≅      1/√2
    \end{minted}
    \caption{Preparing and printing a Bell state.}\label{code:bell}
\end{listing}

\section{Quantum Bits}

\begin{figure}
    \centering
    \hfill
    \subfloat[4-qubits/bits]{\includegraphics[width=.45\linewidth]{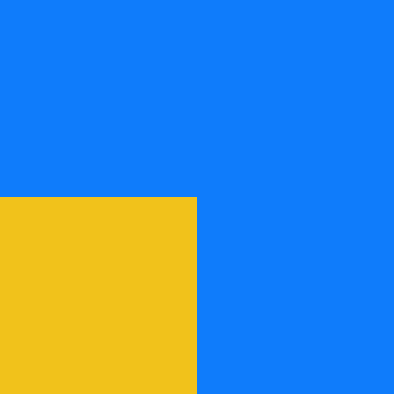}}
    \hfill
    \subfloat[6-qubits/bits]{\includegraphics[width=.45\linewidth]{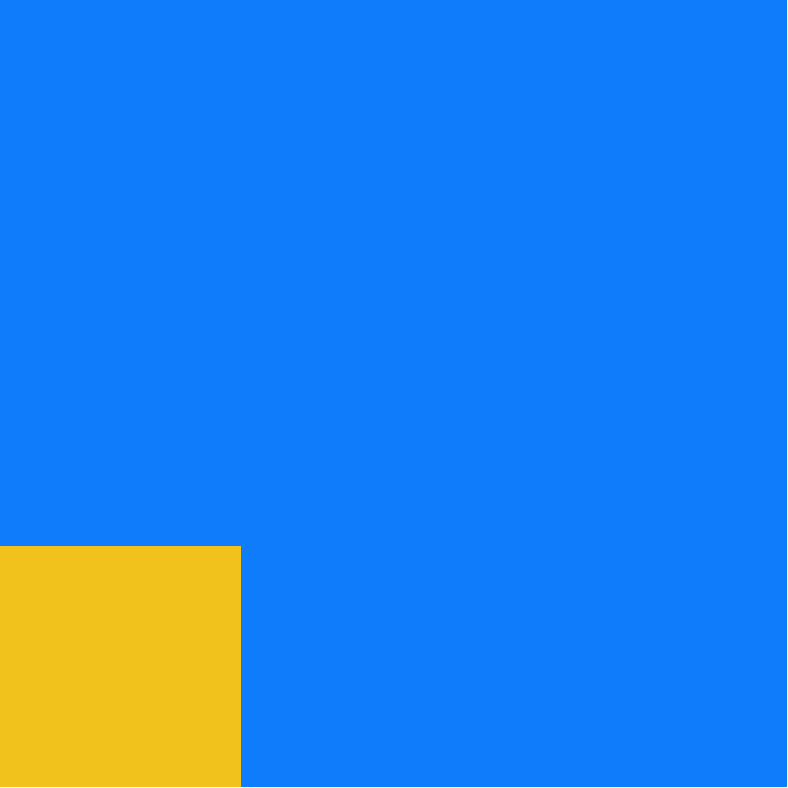}}
    \hfill

    \hfill
    \subfloat[8-qubits/bits]{\includegraphics[width=.45\linewidth]{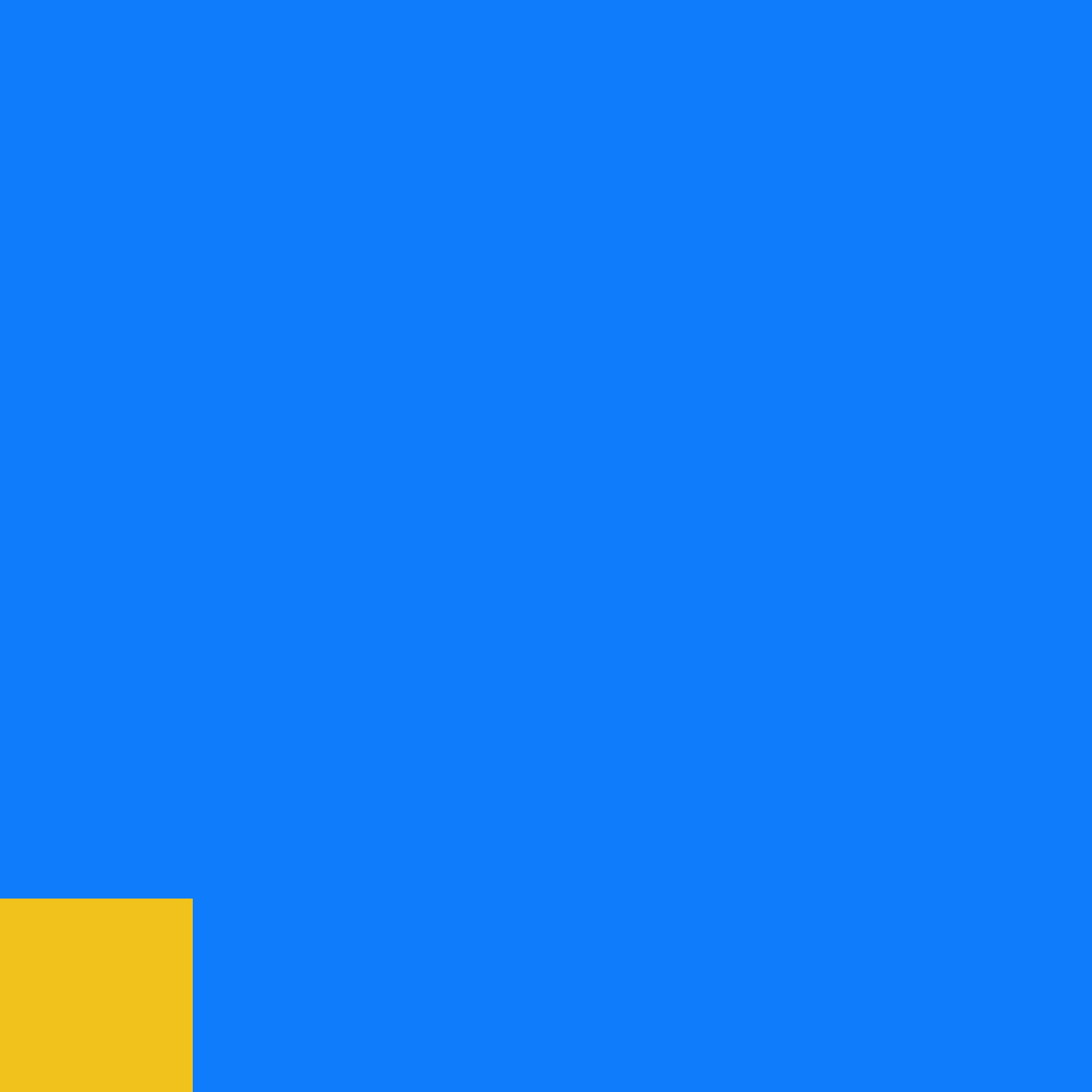}}
    \hfill
    \subfloat[10-qubits/bits]{\includegraphics[width=.45\linewidth]{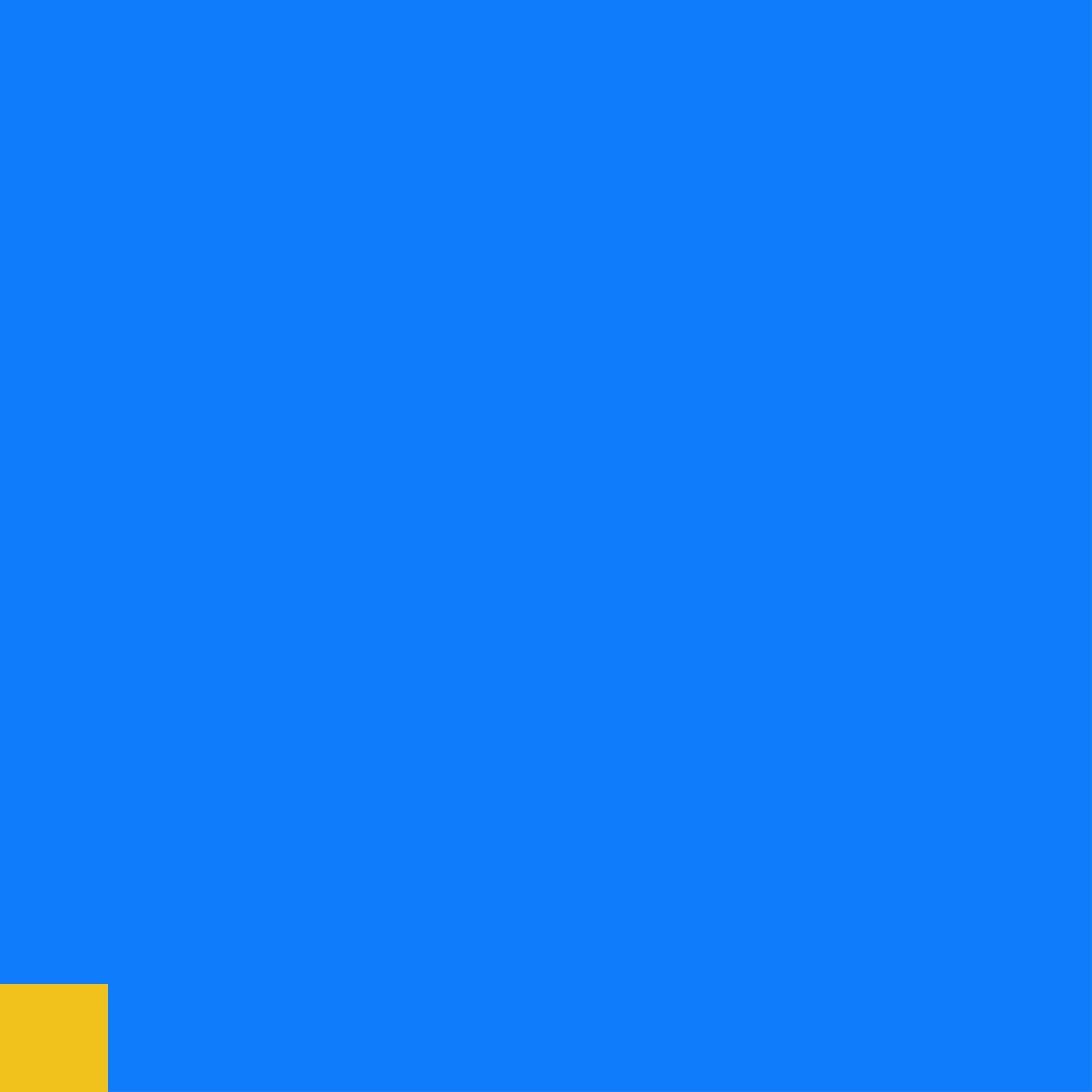}}
    \hfill

    \caption{The total of data $n$-qubits and $n$-bits can hold, respectively, in blue and yellow.}\label{fig:bitqubit}
\end{figure}

Quantum computers store information in quantum bits, aka qubits, the quantum analog for the classical bit. As its classical counterpart, a qubit has two possible values, 0 and 1, or using the Dirac notation, $\ket0$ and $\ket1$, known as the computational basis. A qubit can store classical or quantum information since it can simultaneously be in one or both states, a phenomenon known as quantum superposition. As for classical bits, the number of states in a set of bits grows exponentially with the number of bits, \textit{e.g.}, with 32-bits, one can express a single unsigned integer between 0 and $2^{32}-1$, which is also true for qubits. However, with 32-qubits, one can simultaneously represent all unsigned integers between 0 and $2^{32}-1$. The amount of memory a quantum computer have grows exponentially with the number of qubits, which is equivalent to saying that the number of bits needed to represent a qubit set grows exponentially with the set length. Fig. \ref{fig:bitqubit} geometrically illustrate this exponential memory growth, displaying how much data the same number of qubits and bits can store.

One can represent a qubit in superposition as a linear combination of the computational basis states, with an arbitrary qubit $\ket\psi$ defined as $\alpha\ket0+\beta\ket1$. The numbers $\alpha$ and $\beta$, known as the probability amplitude, weigh the probability of measuring a basis state. We will talk more about measurement later. Those are complex numbers with the restriction of $|\alpha|^2+|\beta|^2 = 1$. Generalizing for more qubits, one can represent $n$-qubits as $\sum_k \alpha_k\ket{k}$, where $\sum_k |\alpha|^2 = 1$.

Quantum mechanics limits how one can modify the state of qubits. For instance, every quantum operation must be reversible except for measurement, meaning there is no loss of information during quantum computation. Reversibility is not necessary for classical computers. We will discuss this in the next section.

Another limitation of quantum computers is the impossibility of copying a quantum state. Nonetheless, quantum information explores this characteristic to offer communication protocols physically secured against eavesdropping.

\subsection{Bloch Sphere}

\begin{figure}[h]
    \centering
    \includegraphics[width=\linewidth]{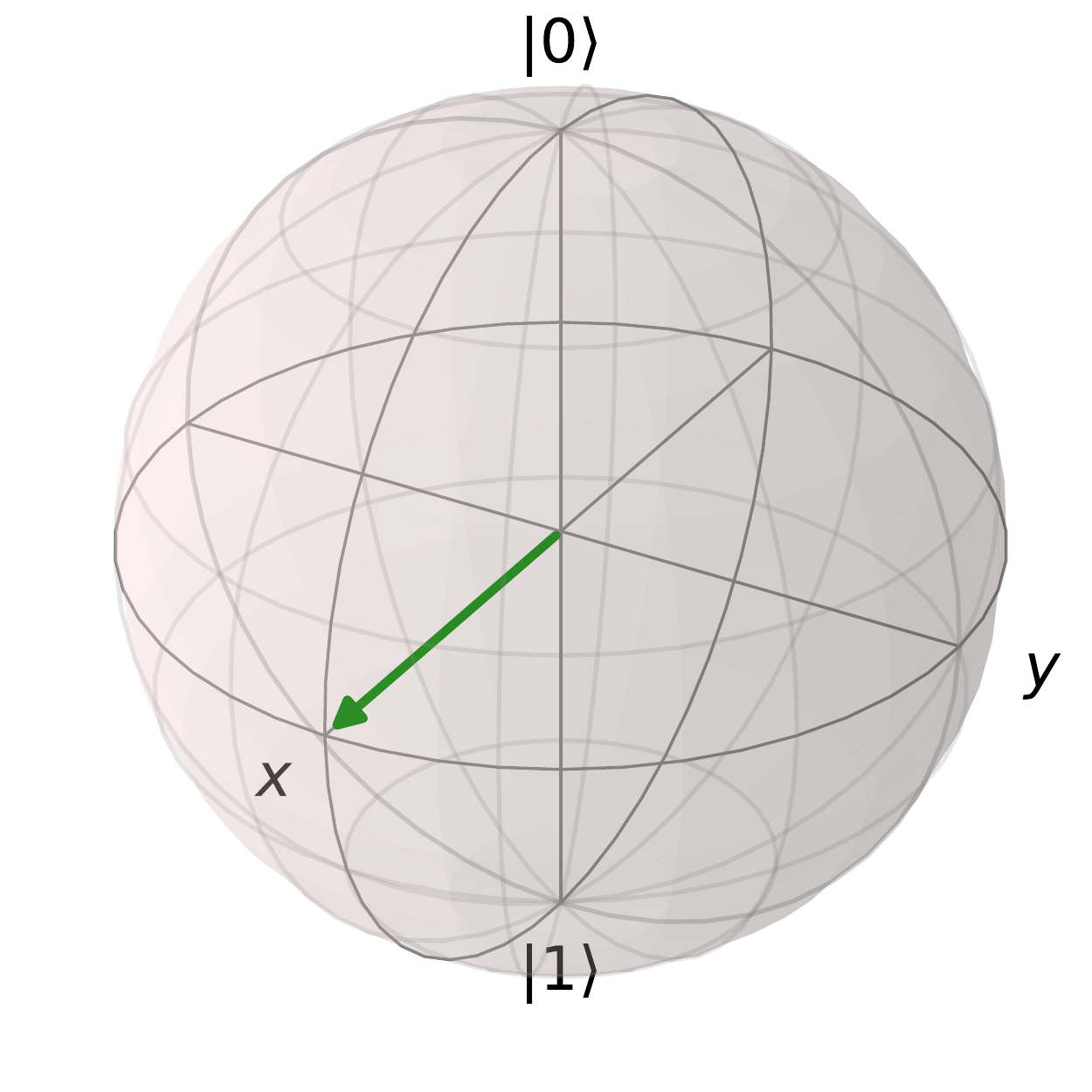}
    \caption{Bloch sphere with a vector in the state $ (1, 0, 0) \equiv \frac{1}{\sqrt{2}}(\ket0+\ket1)$.}\label{fig:bola_plus}
\end{figure}

We can picture every possible single-qubit state in the Bloch sphere. This geometric representation helps visualize the quantum superposition and single-qubit operation. In the Bloch sphere of Fig. \ref{fig:bola_plus}, the basis states $\ket0$ and $\ket1$ are at the north and south poles, which are the vectors (0, 0, 1) and (0, 0, -1), respectively. With the origin point (0, 0, 0) in the middle of the sphere, a qubit is a vector of length of 1.

A noise qubit is in the middle of the sphere, with a length of less than 1, which is a helpful visualization when studying quantum error correction, a research field to mitigate noise in quantum computers.

We can represent any qubit state in the Bloch sphere, but we cannot differentiate qubits that diverge only by a global phase. In other words, we cannot distinguish the state qubit $\alpha\ket0+\beta\ket1$ from $e^{i\theta}(\alpha\ket0+\beta\ket1)$, where  $e^{i\theta}$ is a global phase. For example, state $\frac{i}{\sqrt{2}}\ket0+\frac{i}{\sqrt{2}}\ket1$  is identical to state $\frac{1}{\sqrt{2}}\ket0+\frac{1}{\sqrt{2}}\ket1$ in the Bloch sphere (see Fig. \ref{fig:bola_plus}) since $\frac{i}{\sqrt{2}}\ket0+\frac{i}{\sqrt{2}}\ket1 = i(\frac{1}{\sqrt{2}}\ket0+\frac{1}{\sqrt{2}}\ket1)$, where $i$ is the imaginary number equal to $\sqrt{-1}$.

Representing two or more qubits in the Bloch sphere goes beyond the three dimensions. So, it is not a viable representation.

\subsection{Qubits in Ket}

Since qubits are only available for the quantum computer, the classical computer only stores its references. Ket provides built-in functions that take qubit references to apply a quantum operation on the quantum computer. Ket features a complete set of quantum gates, which one can use to implement any quantum computation. We will discuss more quantum gates in the next section. Ket can verify at classical runtime if a quantum execution is valid, reporting the error before it goes to the quantum computer.

Ket stores qubits in lists of the type \texttt{quant}. The \texttt{quant}'s constructor allocates a given number of qubits, all in the state $\ket0$. For example, in Listing \ref{code:bell}, line 2, Ket allocates two qubits and assigns its reference to variables \texttt{a} and \texttt{b}. Each \texttt{quant} stores only one qubit reference in this example. However, it can hold as many qubit references as needed. From now on, we will refer to the qubits references in a \texttt{quant} only as qubits.

We saw that quantum bits could be in a superposition of classical states, where each state has an associated probability amplitude. We will discuss probability amplitude more when addressing quantum measurement in Section \ref{sec:mea}. We also presented how to allocate and handle qubits reference in Ket. Next, we will see how to apply quantum gates to create superposition and entangle qubits.

\section{Quantum Gates}

There are two ways to modify a quantum state: measuring or applying quantum gates. We will see how measurement works in the next section. Quantum gates are the quantum analog of the logical gates, which are low-level operations that transistors implement to form a classical processor. Table \ref{tab:and_or} shows the truth table of some logical gates. Note that given the output of a logical gate, one cannot precisely determine its input. We call this operation irreversible. In contrast, every quantum gate is reversible, meaning there is no information loss during quantum computation.

\begin{table}[h]
    \centering
    \begin{tabular}{ccc}
        \toprule
        \multicolumn{2}{c}{Input} & Output           \\
        A                         & B      & A AND B \\
        \midrule
        0                         & 0      & 0       \\
        0                         & 1      & 0       \\
        1                         & 0      & 0       \\
        1                         & 1      & 1       \\
        \bottomrule
    \end{tabular}
    \hspace{14pt}
    \begin{tabular}{ccc}
        \toprule
        \multicolumn{2}{c}{Input} & Output          \\
        A                         & B      & A OR B \\
        \midrule
        0                         & 0      & 0      \\
        0                         & 1      & 1      \\
        1                         & 0      & 1      \\
        1                         & 1      & 1      \\
        \bottomrule
    \end{tabular}

    \caption{Truth table for the AND and OR logical gates.}\label{tab:and_or}
\end{table}

In Table \ref{tab:h_cnot}, we present the ``truth table'' of some quantum gates. Note that by the output, one can precisely determine the input state. For example, if the result of a Hadamard gate is $\frac{1}{\sqrt{2}}(\ket{0}-\ket{1})$, we know that the input state is $\ket1$. 

Since qubits work with continuous values,  a three dimensions continuum between $\ket0$ and $\ket1$, there are infinite possible quantum gates. Here, we will present the most standard quantum gates and how to use them in Ket. See Ket's API documentation for a list of available quantum gates and their effects.

\begin{table}[h]
    \centering
    \begin{tabular}{cc}
        \toprule
        Input        & Output                                \\
        $\ket{\psi}$ & $H\ket{\psi}$                         \\
        \midrule
        $\ket{0}$    & $\frac{1}{\sqrt{2}}(\ket{0}+\ket{1})$ \\
        $\ket{1}$    & $\frac{1}{\sqrt{2}}(\ket{0}-\ket{1})$ \\
        \bottomrule
    \end{tabular}
    \hspace{14pt}
    \begin{tabular}{ccc}
        \toprule
        \multicolumn{2}{c}{Input} & Output                                    \\
        $\ket{\psi}$              & $\ket{\varphi}$ & CNOT$\ket{\psi\varphi}$ \\
        \midrule
        $\ket{0}$                 & $\ket{0}$       & $\ket{00}$              \\
        $\ket{0}$                 & $\ket{1}$       & $\ket{01}$              \\
        $\ket{1}$                 & $\ket{0}$       & $\ket{11}$              \\
        $\ket{1}$                 & $\ket{1}$       & $\ket{10}$              \\
        \bottomrule
    \end{tabular}

    \caption{``Truth table'' for the Hadamard ($H$) and CNOT quantum gates.}\label{tab:h_cnot}
\end{table}

As Ket provides tools to print the quantum state, one can experiment with every quantum gate shown here,  changing parameters and concatenating operations to see the effect. To print a quantum state, one needs to create a \texttt{dump} variable passing a \texttt{quant} to its constructor. Listing \ref{code:bell}, line 5, is an example -- note that we used the plus operator to concatenate two \texttt{quant} variables. The method \texttt{show} returns a string with the quantum state that one can print. Call this method at the end of the code since it will invalidate the allocated qubits afterward. We will explain the why in the next section.

\subsection{Pauli Gates}

The three Pauli gates, \texttt{X}, \texttt{Y}, and \texttt{Z}, apply a $\pi$ rotation in axes of the Bloch sphere. Listing \ref{code:plotx} illustrates the application of an \texttt{X} gate in a qubit $\ket0$. Ket makes it easy to plot the state of a single qubit in the Bloch sphere. Try changing the \texttt{X} gate from Listing \ref{code:plotx}, line 3, to another Pauli gate and see the result. The Pauli \texttt{X} gate is also known as the NOT gate since it flips the qubit between the states $\ket0$ and $\ket1$.

\begin{listing}[h]

    \begin{minted}[frame=topline,linenos,xleftmargin=2em]{ket}
from ket import *
q = quant()
X(q) # Pauli X gate
d = dump(q)
d.sphere().show()
\end{minted}
    \includegraphics[width=\linewidth]{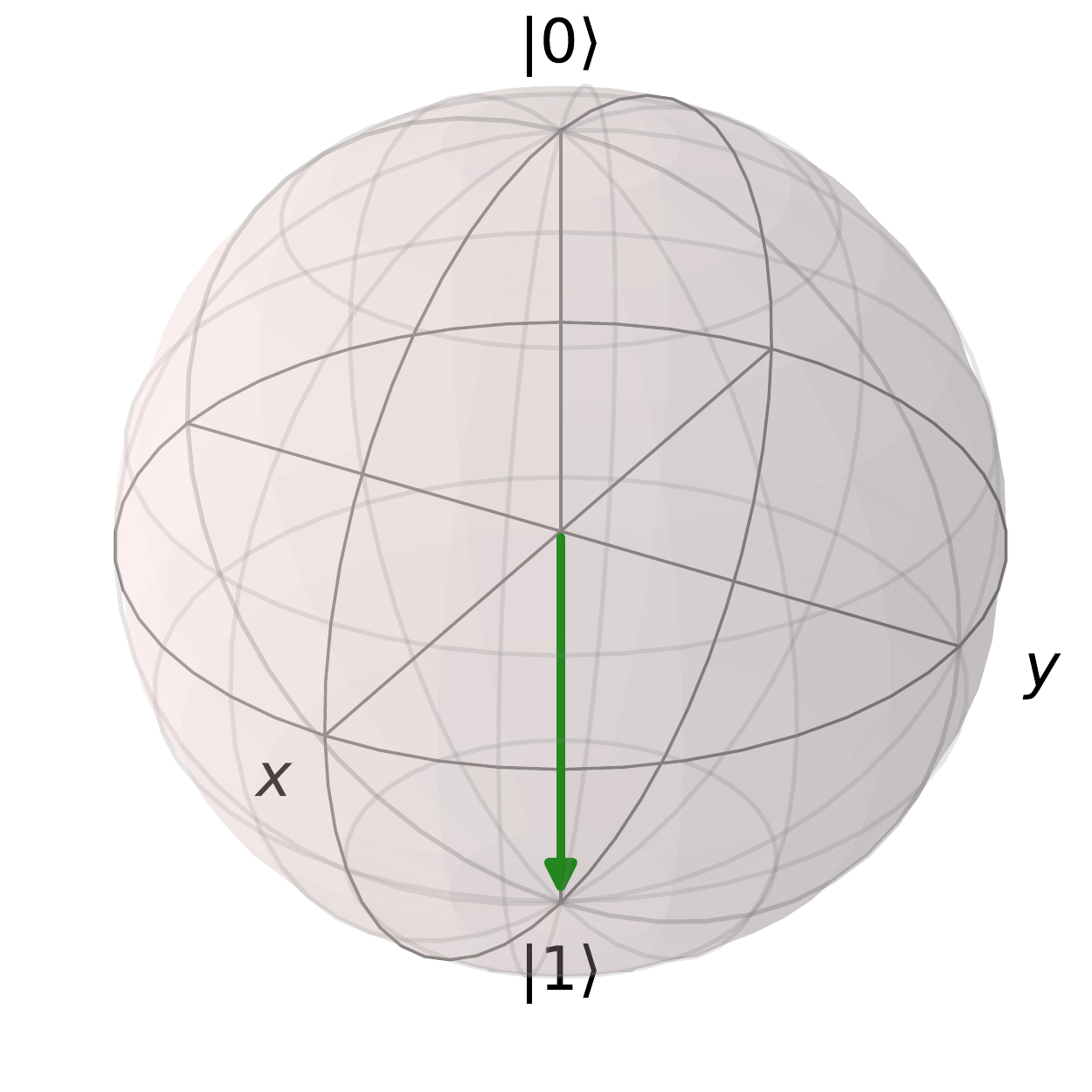}
    \begin{minted}[frame=bottomline,xleftmargin=2em]{text}
    \end{minted}

    \caption{Plotting a Bloch sphere using Ket with a vector on the state $\ket1 \equiv (0, 0, -1)$. For plots, we recommend using Jupyter Notebook.}\label{code:plotx}
\end{listing}

\subsection{Rotation Gates}

Since the Pauli gates do not generate superposition, the states on the Bloch sphere will only jump between $\ket0$ and $\ket1$. One can use an arbitrary rotation gate to put the qubits in a superposition. For that, Ket provides the \texttt{RX}, \texttt{RY}, and \texttt{RZ} gates. For example, rotating a qubit $\frac{\pi}{2}$ around the Y-axes (in Ket: \texttt{RY(pi/2, q)}) generates an equal superposition of the states $\ket0$ and $\ket1$. The rotation gates are enough to express any transformation on a single qubit.

\subsection{Hadamard Gate}

Putting qubits in an equal superposition of the computation basis estates is a typical procedure often used in the preparation stage of quantum algorithms. The Hadamard gate (in Ket: \texttt{H}) does precisely that, equivalent to $\frac{\pi}{2}$ Y-axes rotation followed by a Pauli \texttt{X} gate.

\subsection{Phase Gate}

The Phase gate (in Ket: \texttt{phase($\lambda$, q)}) adds a relative $\lambda$ phase to state $\ket1$. With this gate, we can express some popular gates. By setting lambda to $\pi$, $\frac{\pi}{2}$, and $\frac{\pi}{4}$, we get the gates Pauli \texttt{Z}, \texttt{S}, and \texttt{T}, respectively. This gate is equivalent to the Z-axis rotation gate, diverging only by the global phase.

\subsection{Entanglement and Controlled Gates}

So far, we see how to apply quantum gates to qubits individually but not how to make them interact with each other. For that, we can use a Controlled-Not (CNOT) gate. This gate takes a control and a target qubit, applying a Pauli \texttt{X} gate in the target qubit when the control qubit is on state $\ket1$. The presented single-qubit gates with the CNOT gate are enough to implement any quantum algorithm. See Listing \ref{code:bell}, line 4, for an example of a CNOT application in Ket.

In Table \ref{tab:h_cnot}, we see how the CNOT gate operates in qubits that are not in superposition. However, what happens when the control qubit is in a superposition? When the control qubit is in a superposition of both states $\ket1$ and $\ket0$, the CNOT gate will entangle both qubits to a state where the NOT gate is/is not applied. For example, if the control qubit is in the state $\frac{1}{\sqrt{2}}(\ket0+\ket1)$ and the target qubit is in the state $\ket0$, the state after the quantum gate application will be $\frac{1}{\sqrt{2}}(\ket{00}+\ket{11})$, known as the Bell state. Before the CNOT gate application, the qubits are in a superposition of states $\ket{00}$ and $\ket{10}$. Analyzing each state individually, the CNOT must flip the second bit of state $\ket{10}$ and do nothing on state $\ket{00}$. The resulting quantum state is precisely that, letting the qubits in a superposition of the states $\ket{00}$ and $\ket{11}$.

Indeed, it is possible to add control qubits to any quantum gate application. In Ket, one can use the \texttt{ctrl} function to call a quantum gate with control qubits. Listing \ref{code:ctrlh} shows how to apply a controlled-RY gate.

\begin{listing}[h]
    \begin{minted}{ketcon}
>>> from ket import *
>>> from math import *
>>> a, b = quant(2)
>>> ctrl(RY(pi/2, a), H, b)
>>> print(dump(a+b).show())
|00⟩    (50.00%)
 0.707107           ≅      1/√2
|10⟩    (25.00%)
 0.500000           ≅      1/√4
|11⟩    (25.00%)
 0.500000           ≅      1/√4
    \end{minted}
    \caption{Using the \texttt{ctrl} function to apply a controlled-Hadamard gate in Ket, with \texttt{a} as control and \texttt{b} as the target.}\label{code:ctrlh}
\end{listing}

Ket does not differentiate quantum gate from Python function, letting one add qubits of control to any callable -- with a few restrictions. Ket will raise an error if the controlled call: allocates or frees qubits, measure qubits, or perform classical operations on the quantum computer. In Listing \ref{code:ctrlbell}, we define and call the quantum gate Bell with a control qubit, entangling three qubits. Note that Ket allows nesting controlled calls. Also, the quantum gates implemented in Ket return a reference for the input qubits, allowing the concatenation of quantum gates.

\begin{listing}[h]
    \begin{minted}{ketcon}
>>> def bell(a, b):
...     ctrl(H(a), X, b)
... 
>>> q = quant(3)
>>> ctrl(H(q[0]), bell, q[1], q[2])
>>> print(dump(q).show())
|000⟩    (50.00%)
 0.707107           ≅      1/√2
|100⟩    (25.00%)
 0.500000           ≅      1/√4
|111⟩    (25.00%)
 0.500000           ≅      1/√4
    \end{minted}
    \caption{Calling a controlled-Bell gate in Ket}\label{code:ctrlbell}
\end{listing}

\subsection{Inverse Quantum Gate}

Since quantum gates are reversible operations, many algorithms describe their steps in terms of inverse quantum gates. For example, in one of the last steps of Shor's algorithm, we need to use an inverse quantum Fourier transformation on a list of qubits. If we already have quantum Fourier transformation implemented in Ket, applying an inverse quantum Fourier transformation is trivial. Ket provides the \texttt{adj} function that calls the inverse of a quantum gate. Listing \ref{code:shor} is the quantum subroutine of Shor's algorithm, with the inverse quantum Fourier transformation call using \texttt{adj} in line 7.

\begin{listing}[h]
    \begin{minted}{ket}
def order_find(x, N):
    L = N.bit_length()
    t = 2*L+1
    reg1 = quant(t)
    H(reg1)
    pown(x, reg1, N) # ket.plugins
    adj(qft, reg1) # ket.lib
    return measure(reg1)
    \end{minted}
    \caption{Quantum subroutine for Shor's factoring algorithm.}\label{code:shor}
\end{listing}

Another pattern seen in many quantum algorithms is placing a function \texttt{A} between an operation \texttt{B} and its inverse, with the following form in Ket: \texttt{B(); A(); adj(B)}. To accommodate those constructions, Ket provides the \texttt{with around} statement. Listing \ref{code:grover}, line 2, shows this statement in action, implementing the diffusion operation used in Grover's search quantum algorithm.

\begin{listing}[h]
    \begin{minted}{ket}
def grover_diffusor(s: quant):
    with around(X(H), s): # X(H(s))
        ctrl(s[1:], Z, s[0])
    # adj(X(H), s)
    \end{minted}
    \caption{Grover's diffusor operation in Ket.}\label{code:grover}
\end{listing}

A set of entangled qubits becomes a single entity. We cannot describe its parts separately, and an operation on a single qubit changes the state of all entangled qubits. We can experiment with this behavior using the \texttt{with around} statement. Listing \ref{code:around_bell} uses the \texttt{bell} function defined in Listing \ref{code:ctrlbell} to entangle the two qubits at the beginning of the \texttt{with around} statement, undoing the engagement at the end. Inside the scope initialized in line 2, any operation will reflect on both qubits since they are entangled. In Listing \ref{code:around_bell}, the Pauli \texttt{X} gate applied in the qubit \texttt{a} flips the qubit \texttt{b} at the end. In fact, a Pauli \texttt{X} flip the qubit \texttt{b} independent of the qubit we use it. The same happens for the Pauli \texttt{Z} gate that ends flipping the qubit \texttt{a}. We recommend executing this example, applying different gates to different qubits to see how the operation inside this entangled scope affects both qubits.

\begin{listing}[h]
    \begin{minted}{ketcon}
>>> a, b = quant(2)
>>> with around(bell, a, b):
...     X(a)
... 
>>> print(dump(a+b).show('i1:i1'))
|0⟩|1⟩  (100.00%)
 1.000000               ≅      1/√1
    \end{minted}
    \caption{Using \texttt{with around} to create an entangled code-block.}\label{code:around_bell}
\end{listing}

\section{Quantum Measurement}
\label{sec:mea}

The final result of a quantum execution is classical information. For example, Shor's algorithm uses quantum superposition to find classical data, the prime factor of a number. However, even though qubits can store an exponential amount of data, there is no efficient way to read it.

To get classical information from a qubit, we need to measure it. However, this operation only returns one basis state of the superposition. So, if we have a qubit in the state $\alpha\ket0+\beta\ket1$, a measurement will return 0 or 1 randomly. A measure operation also has the quantum side effect of collapsing the quantum state in the measurement result. For example, if the measurement of the state $\alpha\ket0+\beta\ket1$ returns 1, the quantum state after the measure will be $\ket1$, and any measure operation afterward will return 1.

We need to measure a qubit several times to gather enough information to reconstruct a quantum state in a classical computer. Although, since the measurement collapses the quantum state, we also need to prepare the quantum state several times. For example, considering the state $\sum_{k=0}^{2^n-1} \alpha_k\ket{k}$ with $n$ qubits in a superposition of $2^n$ basis states, we need to prepare and measure the state more than $2^n$ times to know all the basis states. So, for a quantum algorithm to have an advantage over its classical counterpart, it must be possible to extract the result of the quantum superposition using a small number of executions and measurements.

The squared module of the \emph{probability amplitude} is the probability of measuring a basis state. For example, the chance of measuring 0 or 1 of the state $\alpha\ket0+\beta\ket1$ is $|\alpha|^2$ and $|\beta|^2$, respectively. We can generalize this for an $n$-qubit state  $\sum_{k=0}^{2^n-1} \alpha_k\ket{k}$, where the probability of measuring $k$ is $|\alpha_k|^2$.

Ket provides the \texttt{measure} function for measuring a list of qubits. This function collapses the quantum state and returns an integer representing the computational basis state measured. For example, measuring the Bell state $\frac{1}{\sqrt{2}}(\ket{00}+\ket{11})$ will return the integer 0 or 3. Ket allows using the collapsed qubits after measurements. However, most quantum computers today invalidate the qubit after a measure because of technical limitations. Before demonstrating how to use the \texttt{measure} function in Ket, we need to understand how Ket handles the quantum execution to get the result from the quantum computer.

\subsection{Ket Runtime Architecture}

\begin{figure}[h]
    \centering
    \includegraphics[width=\linewidth]{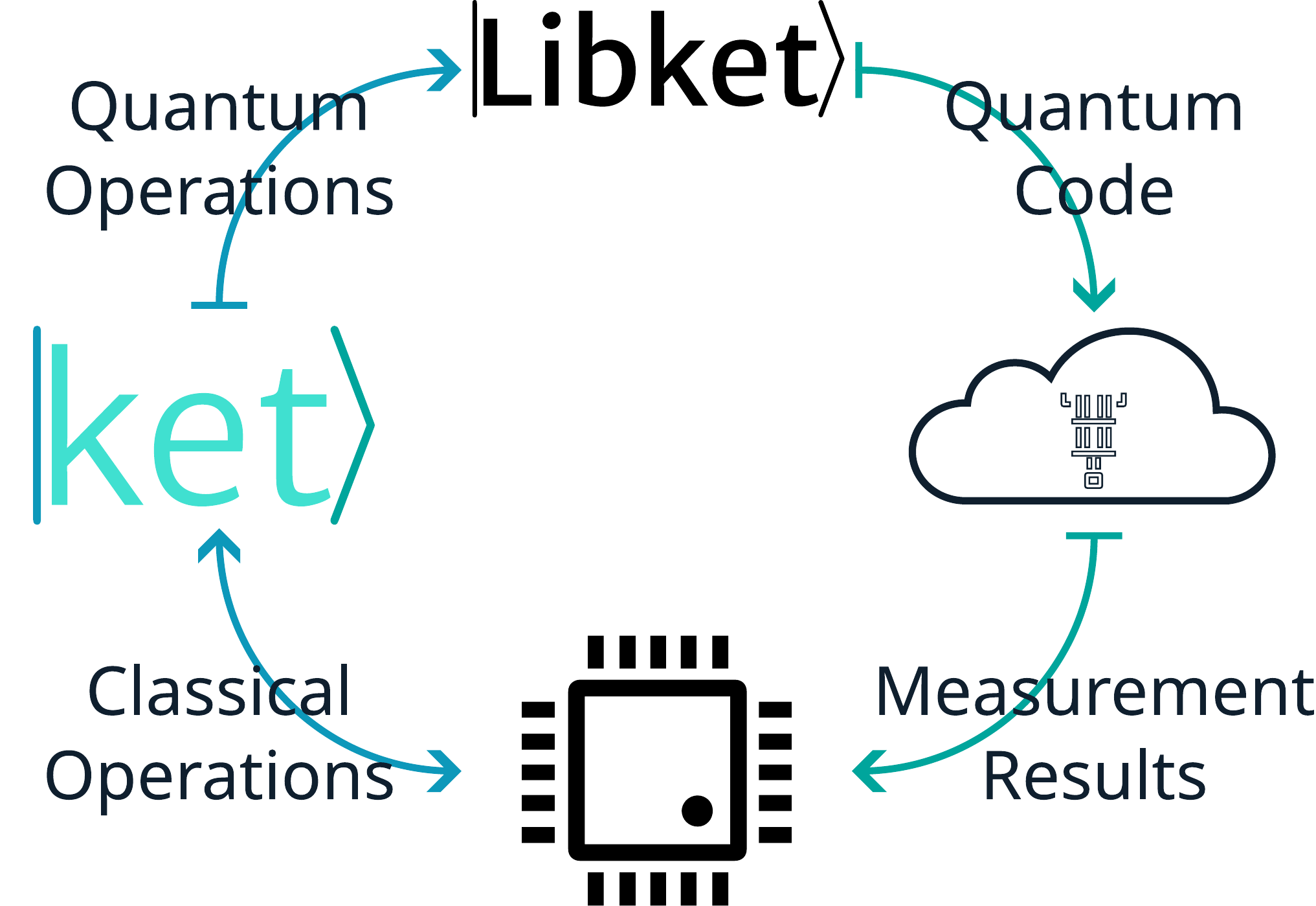}
    \caption{Ket classical-quantum execution loop with classical runtime composed by the Ket application (left), the runtime library Libket (top), and the CPU (bottom), and quantum runtime represented by the quantum computer (right).}\label{fig:runtime}
\end{figure}

Since a quantum computer works as a coprocessor for a classical computer, usually accessible only through the cloud, Ket splits its execution into classical runtime and quantum runtime. The classical runtime is in charge of executing classical instruction, creating the quantum code, and processing the quantum result. In contrast, the quantum runtime is responsible for running the quantum code. Ket intercalates between classical and quantum runtime in a loop, as illustrated in Fig. \ref{fig:runtime}, with the following phases:

\begin{enumerate}
    \item At classical runtime, Ket issues (most of) the classical operation to execute on the local CPU and passes the quantum operation to its runtime library, called Libket;
    \item Libket, still at classical runtime, uses the received quantum operation to build quantum code with all the instructions needed for the quantum execution;
    \item When the classical computer needs a quantum measurement result, Libket sends the quantum code for execution, starting the  quantum runtime;
    \item At quantum runtime, the quantum computer or simulator executes the quantum code and returns the measurement result for the classical computer -- There is no iteration between classical and quantum computers in the middle of the quantum runtime, so the classical computer cannot require partial information about a quantum execution nor send more instructions during it;
    \item The quantum runtime ends when the classical computer receives the quantum measurement result, starting the classical runtime again to process end interpret the data;
    \item The loop starts again, enabling the execution of an interactive classical-quantum application.
\end{enumerate}

In this architecture, if Ket returns the measurement result right after, the quantum context will be lost, forbidding further operations on the allocated qubits. So, instead, Ket returns a \texttt{future} variable as a promise that the measurement result will be available for the classical computer when needed. To get the measurement result from the quantum computer, one needs to read the \texttt{.value} attribute of the returned \texttt{future} variable. For example, Listing \ref{code:bell_mea} shows the preparation and measurement of qubits in the Bell state. On the other hand, Listing \ref{code:bell_err} raises an error in line 5 because we already read measurement results from line 4, which invalids the allocated qubits, forbidding us to proceed with a new measure.

\begin{listing}[h]
    \centering
    \begin{minted}{ketcon}
>>> from ket import *
>>> a, b = quant(2)
>>> cnot(H(a), b)
>>> m_a = measure(a)
>>> m_b = measure(b)
>>> print(m_a.value, m_b.value)
1 1
    \end{minted}
    \caption{Preparing and measuring a Bell state.}\label{code:bell_mea}
\end{listing}

\begin{listing}[h]
    \centering
    \begin{minted}[highlightlines={5},highlightcolor=orange]{ketcon}
>>> from ket import *
>>> a, b = quant(2)
>>> cnot(H(a), b)
>>> m_a = measure(a).value
>>> m_b = measure(b).value
    \end{minted}
    \caption{This code raises an error in line 5 because the qubit \texttt{a} is no longer valid after line 4.}\label{code:bell_err}
\end{listing}

Reading the result of quantum measurement tells Libket that the classical computer needs the information, triggering a quantum execution. Ket forbids the communication between classical and quantum computers during quantum execution, with the premise that a quantum computer needs to execute as fast as possible to mitigate decoherence and quantum errors. With this restriction, two quantum executions cannot interfere with each other. So Ket invalidates all the allocated qubits after the quantum execution, which is the motivation behind the error raised in Listing \ref{code:bell_err}. After a quantum execution, one can instantiate new qubits to start a new quantum execution. Variables of the type \texttt{dump} follow the same rules, so reading any quantum data from it will trigger the quantum execution.

As a qubit in Ket is a reference for a quantum value in the quantum computer, a \texttt{future} variable is a reference for a classical value in the quantum computer. We can use \texttt{future} variables to evaluate classical exceptions and perform control flow on the quantum computer. The following subsection illustrates this with a Ket implementation of the quantum teleportation protocol.

\subsection{Measuring Entangled Qubits}

As we mentioned, an entangling qubit set behaves as a single entity, with any operation in one qubit affecting all qubits, including measurement. Considering that we have two qubits $\ket{\psi} = \frac{1}{\sqrt{2}}(\ket{00}+\ket{11})$ on the Bell state, we know that the probability of measuring 1 on the first qubits is 50\%, but what is its quantum state after measuring it? If we measure 1 on the first qubit, the qubits will collapse on the basis states with 1 on the first qubit, which is the state $\ket{11}$. Collapsing the qubits to $\ket{11}$ implies that a measurement on the second qubits will always return 1. The same is true for measuring 0 on the first qubit, which results in measures of the second qubit always returning 0. You can see this in Listing \ref{code:bell_mea}, where the measurements of both qubits will always return the same result, \texttt{0 0} or \texttt{1 1}.

Measuring one qubit will return information from all the qubits it is entangled. Furthermore, this characteristic is independent of the distance. For example, considering that we have two entangled qubits in the Bell state, one on Earth and one on Mars, if, at the same time, we measure both qubits, the result will always be equal, 0 on Earth and 0 on Mars or 1 on Earth and 1 on Mars. Although both qubits interact faster than the speed of light to collapse in the same state, this does not make faster-than-light communication possible. Nevertheless, many quantum communication protocols use this characteristic, \textit{e.g.}, the quantum teleportation protocol.

Listing \ref{code:tele} presents the implementation of the quantum teleportation protocol in Ket. This protocol teleports the information of a qubit from Alice to Bob using a shared pair of entangled qubits (lines 19-21). The protocol starts with Alice entangling the qubit that she what to teleport with her qubit from the entangled pair and measuring both qubits (lines 22-25), collapsing Bob's qubit into a quantum state close to Alice's. To transform Bob's qubit into the identical qubit from Alice, he can use the measurement results from Alice's qubits (line 26). This protocol uses classical and quantum channels to transmit the bits and the entangled qubits.

\begin{listing}[h]
    \centering
    \begin{minted}{ket}        
def entangle(a, b):
    return cnot(H(a), b)

def teleport(quantum_message, 
             entangled_qubit):
    adj(entangle,
        quantum_message, 
        entangled_qubit)
    m0 = measure(entangled_qubit)
    m1 = measure(quantum_message)
    return m0, m1

def decode(classical_message, qubit):
    if classical_message[0] == 1:
        X(qubit)
    if classical_message[1] == 1:
        Z(qubit)

alice_message = phase(pi/4, H(quant()))
bell = entangle(*quant(2))
alice_qubit, bob_qubit = bell
classical_message = teleport(
    quantum_message=alice_message,
    entangled_qubit=alice_qubit
)
decode(classical_message, bob_qubit)
print(measure(bob_qubit).value)
\end{minted}
    \caption{Quantum teleportation protocol in Ket.}\label{code:tele}
\end{listing}

Note that in Listing \ref{code:tele}, function \texttt{decode}, we do not read the \texttt{.value} attribute of any \texttt{future} variable in the list \texttt{classical\_message}. So the expressions and statements in lines 14 to 17 can only be evaluated by the quantum computer since the classical computer do not have those values. Ket only sends the quantum code for execution when we print the result of Bob's qubit measurement at the end of the code.

For the correct evaluation of the \texttt{if} statements in Listing \ref{code:tele}, swap the Python interpreter with Ket's interpreter or add the decorator \texttt{@ket.code\_ket} on the \texttt{decode} function\footnote{The \texttt{decode} function may not work correctly inside a Jupyter Notebook. We recommend executing this code from a \texttt{.py} or \texttt{.ket} file.}. We recommend executing Listing \ref{code:tele} on your computer, changing Alice's qubit state, and seeing how it arrives in Bob's qubit. It is also possible to dump the quantum state before the \texttt{if} statements to see how Alice's qubits interfere with Bob's.

\section{Quantum Computer vs. Simulator}

Ket packages a quantum simulator so one can execute a quantum application out of the box. Ket sends the quantum code for the Ket Bitwise Simulator (KBW) by default. However, one can define where the quantum code will execute. The setup necessary for that depends on the quantum execution target. Although, it will unlikely require changing the quantum program logic. Quantuloop offers high-performance quantum computer simulators for Ket to speed up quantum development. Also, Quantuloop is developing interfaces to execute Ket applications on quantum computers.

Quantum simulators do not substitute quantum computers, and neither do quantum computers make quantum simulators obsolete. Even with quantum simulation being a computationally-intensive problem, laptops can easily simulate around twenty qubits. Although we cannot do meaningful work with so few qubits, it is enough to execute small instances of most quantum algorithms. So, in the same sense that we use MIPS simulators in computer organization courses, we can use quantum simulators in quantum computing education.

We can simulate more qubits when solving real-world problems than is possible on quantum computers due to quantum errors. So, today, we can use quantum simulators to validate quantum applications, getting ready for when large-scale fault-tolerant quantum computers become available. Since Ket is a general-purpose quantum programming language, one can implement and validate a quantum application today and execute it on a quantum computer in the future.

Since quantum computers have physical constraints that make it impossible to look at quantum superposition, we cannot debug a quantum application as we classically do. For example, a quantum computer cannot efficiently return the quantum state in a \texttt{dump} variable as a quantum simulator does. Nonetheless, using \texttt{dump} variables with a quantum simulator help understand how a quantum application executes, facilitating the test of quantum programs.

\section{Final Considerations}

As this is a gentle introduction to quantum programming, we abstract most of the mathematical formalism behind quantum computing. For instance, we can define every quantum operation with linear algebra, where qubits are vectors; and quantum gates are matrixes. We encourage anyone new to the field to read the postulates of quantum mechanics enumerated by \cite{nielsen_chuang_2010} and understand the linear algebra of quantum computing. This knowledge will help the comprehension of the inner works of quantum algorithms.

For those who know quantum computing, quantum programming with Ket will not be a drag to develop a classical-quantum application. We only cover the basic functionalities of Ket here, yet it is enough for most quantum developments. See the project's website for more information on Ket and the paper \cite{10.1145/3474224} for a deeper look at the particularities of quantum programming.

Quantum computing is not science fiction anymore, we already have quantum computers today, and soon they will accelerate several tasks, leading to a new era of innovation, discoveries, and evolution, and you can be part of it. The development of quantum computers is well underway, and now, we need to start integrating them into our processes to get the benefits of quantum computing.

\bibliographystyle{IEEEtran}
\bibliography{main}

\begin{thebibliography}{1}
\providecommand{\url}[1]{#1}
\csname url@samestyle\endcsname
\providecommand{\newblock}{\relax}
\providecommand{\bibinfo}[2]{#2}
\providecommand{\BIBentrySTDinterwordspacing}{\spaceskip=0pt\relax}
\providecommand{\BIBentryALTinterwordstretchfactor}{4}
\providecommand{\BIBentryALTinterwordspacing}{\spaceskip=\fontdimen2\font plus
\BIBentryALTinterwordstretchfactor\fontdimen3\font minus
  \fontdimen4\font\relax}
\providecommand{\BIBforeignlanguage}[2]{{%
\expandafter\ifx\csname l@#1\endcsname\relax
\typeout{** WARNING: IEEEtran.bst: No hyphenation pattern has been}%
\typeout{** loaded for the language `#1'. Using the pattern for}%
\typeout{** the default language instead.}%
\else
\language=\csname l@#1\endcsname
\fi
#2}}
\providecommand{\BIBdecl}{\relax}
\BIBdecl

\bibitem{nielsen_chuang_2010}
M.~A. Nielsen and I.~L. Chuang, \emph{Quantum Computation and Quantum
  Information: 10th Anniversary Edition}.\hskip 1em plus 0.5em minus
  0.4em\relax Cambridge University Press, 2010.

\bibitem{10.1145/3474224}
\BIBentryALTinterwordspacing
E.~C.~R. da~Rosa and R.~de~Santiago, ``Ket quantum programming,'' \emph{J.
  Emerg. Technol. Comput. Syst.}, vol.~18, no.~1, oct 2021. [Online].
  Available: \url{https://doi.org/10.1145/3474224}
\BIBentrySTDinterwordspacing

\end{thebibliography}

\end{document}